\documentclass[letterpaper]{article} 
\usepackage{aaai25}  
\usepackage{times}  
\usepackage{helvet}  
\usepackage{courier}  
\usepackage[hyphens]{url}  
\usepackage{graphicx} 
\usepackage{natbib}  
\usepackage{caption} 
\usepackage{newfloat}
\usepackage{listings}

\usepackage{comment}  
\usepackage{algorithm, algpseudocode} 
\usepackage{amsmath,amssymb} 
\usepackage[subrefformat=parens]{subcaption} 
\usepackage{here} 
\DeclareCaptionStyle{ruled}{labelfont=normalfont,labelsep=colon,strut=off} 
\floatstyle{ruled}
\newfloat{listing}{tb}{lst}{}
\floatname{listing}{Listing}
\usepackage{amsmath,amssymb}
\usepackage{pifont}

\title{Thoughts on Objectives of Sparse and Hierarchical Masked Image Model}
\author {
    Asahi Miyazaki \textsuperscript{\rm 1, \rm a},
    Tsuyoshi Okita\textsuperscript{\rm 1, \rm b}
}
\affiliations {
    \textsuperscript{\rm 1}Kyushu Institute of Technology\\
    \textsuperscript{\rm a}miyazaki.asahi676@mail.kyutech.jp,
     
    \textsuperscript{\rm b}tsuyoshi@ai.kyutech.ac.jp
}

\usepackage{bibentry}

\begin{document}
\maketitle

\begin{abstract}

Masked image modeling is one of the most poplular objectives of training. 
Recently, the SparK model has been proposed with superior performance among self-supervised learning models.
This paper proposes a new mask pattern for this SparK model, proposing it as the Mesh Mask-ed SparK model. We report the effect of the mask pattern used for image masking in pre-training on performance.
\end{abstract}

\section{Introduction}

One of the major frameworks driving LLM and generative AI in recent years is generative models. In this paper, we focus on transformers~\cite{Vaswani2017}, one of the generative models. Transformers were originally architectures for languages. After the transformer for vision was proposed~\cite{vitpaper}, a number of architectures rooted in the vision transformer were presented. The vision transformer employs a two-stage architecture called self-supervised learning, where in the first stage, a large-scale pre-training model is constructed using large but unlabeled training data, and in the second stage, this large-scale pre-training model is used to solve downstream tasks.

The learning objective in vision transformer pre-training is Masked Image Modeling (MIM). Masked Image Modeling is a pre-training method that first masks (hides) a part of the image and then asks the model to guess the masked part. This is an application of BERT~\cite{devlin2018bert}'s masked image model to images. The term “masked image model” itself has been used since BEiT~\cite{mimpaper}. Other learning objectives used in vision transformers include contrast learning used in SimCLR~\cite{simclrpaper}.

SparK, a model derived from such a mask image model, was presented in 2023 by Tian et al~\cite{sparkpaper}. It introduces learning objectives that incorporate sparsity and hierarchy into the mask image model. Sparsity aims to make the connections in the network sparse, while network hierarchy overlaps with the motivation often used in image recognition to extract features at various scales from the image. It is reported to achieve better performance than other methods on the tasks of image classification, object detection, and instance segmentation.

MIM first divides the image into a grid. Each element after segmentation is called a patch. Next, a portion of all patches
Then, all the patches are partially masked, which means that the patches are hidden. BEiT uses block-wise masking as the algorithm for determining which patches to mask. The algorithm selects multiple rectangular patches from the image divided into a grid and masks those areas. The algorithm is described in Section \ref{OurMethod}. The set of patches partially masked by this masking process is fed to the model, which learns to recover the masked patches based on the unmasked patches.







\section{Related Work} \label{関連研究}

\subsection{SparK}

SparK is a self-supervised learning model that uses MIM for pretext tasks. SparK achieves high performance by hierarchizing the network structure to extract features at various scales, and by incorporating sparsity to improve generalization performance and prevent loss of mask patterns.

\subsection{Masking Method of MIM}

SparK pre-training uses Masked Image Modeling (MIM), a method of self-supervised learning that masks an image and performs pretext reconstruction of the masked areas. There are several ways to generate the masks used in this process. For example, SimMIM~\cite{xie2022simmim}, a simple MIM model, uses three masking methods: square, block-wise, and random. Block-wise masks were originally proposed by BEiT~\cite{mimpaper}, while random masks are used by SparK. In the SimMIM paper, the best performance was achieved using the random mask, with a correct response rate of 83.0\%, when the patch size was 32 × 32 and the mask accounted for 50\% of the total patches.

\section{Experiment Setting of Self-Supervised Learning}\label{basic-information}








\subsection{Datasets}

The dataset used in this paper consists of a total of 12094 CT brain images. The images are horizontal slices of the human brain, 512px x 512px in size. All images are monochrome images with only one channel. All slices were randomly divided into four categories: pre-training data, training data for downstream, validation data, and test data as shown in Table \ref{table:DatasetDivision}. The training, validation, and test data have correct labels, but the pre-training data do not. Brain images with tumors were considered positive, and those without tumors were considered negative. The entire dataset is first divided into pre-training data and downstream data for each hospital. Therefore, the pre-training data and downstream data do not contain slices of the same patient at the same time. All slices in the downstream data are then mixed and divided into training, validation, and test slices. The pre-training data contains data for 781 patients, while the downstream data contains data for 76 patients.

\begin{table}[h]
 \caption{Dataset Division}
 \label{table:DatasetDivision}
 \centering
  \begin{tabular}{cl}
   \hline
   Dataset for & Number \\
   \hline \hline
        Pretraining & 10313 \\
        Training & 1258(Negative) + 166(Positive) = 1424 \\
        Evaluation & 152(Negative) + 26(Positive) = 178 \\
        Test & 158(Negative) + 21(Positive) = 179 \\
   \hline
  \end{tabular}
\end{table}

\subsection{Metrics for Performance Evaluation}

The dataset used in this experiment is unbalanced, with about eight times as many negative images as positive images. In such an unbalanced dataset, using only the percentage of correct responses is problematic when evaluating the model. For example, a model that judges all test data negative has a correct response rate of approximately 88\%, even though it does not serve the purpose of detecting brain tumors at all. Therefore, using the percentage of correct responses as a performance indicator is inappropriate for this experiment. To correctly determine whether brain tumors are detected, it is necessary to know the percentage of images that are actually judged positive out of those that should be judged positive. Therefore, the reproducibility is used as an evaluation index. Furthermore, since this indicator alone cannot determine the false positive rate, the fit rate is also used as an evaluation indicator. The F1 score is used as an overall evaluation of the model's accuracy. The recall, precision, and F1 scores are calculated as follows, where $TP$ is the number of true positives, $FP$ is the number of false positives, $TN$ is the number of true negatives, and $FN$ is the number of false negatives.

\begin{equation}\label{equ-recall}
recall = \frac{TP}{TP + FN}
\end{equation}

\begin{equation}\label{equ-precision}
precision = \frac{TP}{TP + FP}
\end{equation}

\begin{equation}\label{equ-f1}
F_1 = 2 \cdot \frac{precision \cdot recall}{precision + recall}
\end{equation}

\subsection{Environment}\label{exam-environment}

In this experiment, we used MMPreTrain~\cite{mmpretrainpage}, a PyTorch~\cite{pytorchpage}-based pre-training tool. It allows users to write Python code to configure settings related to training, including how to load and use datasets, configure models, configure hyperparameters, configure data augmentation, and configure the training scheduler. The version used in this experiment is version 1.1.0, released on October 12, 2023. MMPreTrain ships with several default configuration files for training, including SimCLR, MFF, and SparK, which were used as the basis for this experiment. The main modifications were made to the parameters for reading images from the dataset and to the data augmentation settings, while the SparK settings were modified to handle image masking in the pre-training. The machine used for the experiments is equipped with an NVIDIA GeForce RTX 3090 GPU, an AMD Ryzen 5 5600X CPU, and 32 GB memory.

\subsection{Data Augmentation}\label{downstreamtask-data-aug}

The total number of brain image datasets used in this experiment is 12094. On the other hand, the ImageNet dataset contains more than one million images~\cite{imagenet-paper}, so the dataset used in this experiment is not large. When the dataset is small, the model is overfitted to the training data, resulting in poor performance on non-training data~\cite{shorten2019survey}. Data augmentation is a method of increasing the diversity of a dataset by applying some transformation to the data.

However, even if a data augmentation is effective for one dataset, it is not necessarily effective for another dataset. For example, a character loses its original meaning if its left and right sides are reversed. Therefore, data augmentation that randomly reverses the left and right sides of an image is inappropriate for the character recognition task. In the experiments at Section \ref{compare-ssls}, four data augmentations, RandomResizedCrop, RandomFlip, Mixup~\cite{zhang2018mixup}, and CutMix~\cite{cutmix}, are used on a brain image dataset. We investigate the effectiveness of these four data augmentations on brain image datasets.

RandomResizedCrop randomly crops a portion of the image. The aspect ratio of the cropped area is also random. The cropped area is enlarged to a certain size. In this experiment, the size of the cropped area is randomly selected between 8\% and 100\% of the original image, and the aspect ratio is randomly selected between $\frac{3}{4}$ and $\frac{4}{3}$. RandomFlip flips the original image left or right with a probability of 50\%. CutMix and Mixup combine two images of different classes, but in different ways. CutMix pastes the other image directly into the image, while Mixup combines the two images pixel by pixel. The pixel-by-pixel merging method is shown in the equation \ref{equ-mixup-image-mix}. The pixel values of the two images are $x_a$ and $x_b$, respectively, the pixel values of the combined image are $x_c$, and the mixing ratio is $\lambda$.

\begin{equation}\label{equ-mixup-image-mix}
x_c = x_a\lambda + x_b(1-\lambda)
\end{equation}

In addition, CutMix and Mixup also compose labels. The composition method is as shown in the formula \ref{equ-mixup-cutmix-label-mix}. The labels of the two images are $y_a$ and $y_b$, respectively, and the label after composition is $x_c$.

\begin{equation}\label{equ-mixup-cutmix-label-mix}
y_c = y_a\lambda + y_b(1-\lambda)
\end{equation}

In CutMix and Mixup, the mixing ratio $\lambda$ of images and labels is randomly determined by the beta distribution $Beta(\alpha,\alpha)$.

\subsection{Class Weighting}\label{class-label-subsection}

As shown in Table \ref{table:DatasetDivision}, there is an imbalance in the training data, with negative images outnumbering positive images by a factor of about 8. This is expected to cause a problem that reduces the sensitivity of the model to positive images. Some MMPreTrain settings use the label smoothing cross-entropy error as the loss function, but this function does not allow for class weighting, so it was changed to the cross-entropy error. entropy error. The class weights are the inverse of the percentage of images of that class in the training data. In this way, the more images a class contains in the dataset, the smaller its weight will be, while the fewer images it contains in the dataset, the larger its weight will be. Let $ n_0 $ be the number of negatives in the training data and $ n_1 $ be the number of positives in the training data, the weights $ w_0 $ for the negatives and $ w_1 $ for the positives are defined by the formula \ref{equ-class-weight} respectively.

\begin{equation}\label{equ-class-weight}
w_0 = \frac{n_0 + n_1}{n_0}, w_1 = \frac{n_0 + n_1}{n_1}
\end{equation}

\section{Comparison of Self-Supervised Learning Methods}\label{compare-ssls}

In this section, we explain the features of SparK by comparing SparK with SimCLR and MFF. First, the architectures and training methods of SimCLR, MFF, and SparK are described, and finally, the performance of SimCLR, MFF, and SparK are compared. The experiments comparing SimCLR, MFF, and SparK are based on the standard settings provided by MMPreTrain, and we created our own settings modified for use with brain image data sets.

\subsubsection{SimCLR}

SimCLR~\cite{simclrpaper} is a method for learning visual representations using contrastive learning, a method of self-supervised learning. The pre-training of contrastive learning involves a pretext task in which data are compared against each other. Specifically, the model is trained to produce outputs that are closer in distance to data belonging to the same class (positive examples) and farther apart from data belonging to different classes (negative examples). In SimCLR, pre-training is performed using contrastive learning, and supervised learning is performed using the pre-trained model in the downstream task.

We now describe the specific algorithm of SimCLR. As explained earlier, contrast learning requires the definition of positive and negative examples, and SimCLR uses two images of the same image with randomly applied data augmentation as positive examples. SimCLR's model learns to make the outputs of the positive examples closer together and the outputs of the negative examples farther apart.

Next, the details of the data augmentation to create a positive example are described. A mini-batch is created from the dataset and copied into two copies. Each copy is called a view. Then, a data augmentation is randomly applied to each view. The data augmentations are The data augmentations consist of four in total, using RandomResizedCrop, RandomFlip, ColorJitter, and GaussianBlur in that order. RandomResizedCrop randomly crops the image from 20\% to 100\% and randomly changes the aspect ratio between $\frac{3}{4}$ and $\frac{4}{3}$. Finally, the image size is changed to 224 x 224, the input size to the model; RandomFlip flips the image left and right 50\% of the time; ColorJitter randomly changes the brightness and contrast of the image 80\% of the time. In this experiment, we used the standard settings for ColorJitter in MMPreTrain. Note that since all images used in this experiment are one-channel monochrome images, saturation and hue were not changed. GaussianBlur blurs the image with a probability of 50\%. After these operations, the final result is two views containing images of size 224x224. These two views are input to the model.

In MMPreTrain, the model structure is divided into three main modules: backbone, neck, and head. The backbone is the first of these three modules to receive input. In SimCLR, a ResNet~\cite{he2015deep} with a depth of 50 is used. The input image size is 224 × 224, although it is not explicitly set in the configuration. Both of the previous two views are input to the backbone, but they do not intersect, and the output is independent for each. That is, if the two views are $x_1$ and $x_2$ and the outputs are $y_1$ and $y_2$, respectively, then there is a one-to-one correspondence between one view and one output as in the following equation.

\begin{equation}\label{equ-simclr-backbone}
y_1 = ResNet(x_1), y_2 = ResNet(x_2)
\end{equation}

NonLinearNeck is used in the neck. This corresponds to the all-coupling layer called the projection head in the SimCLR paper ~\cite{simclrpaper}. In this layer, as in the backbone, there is a one-to-one correspondence between one view and one output.

The head uses ContrastiveHead, which provides a function to compute the loss from the output from the neck. The loss calculation is performed for both positive and negative examples. That is, if the source images in the two views are the same, the closer the output from the neck, the smaller the loss; if the source images in the two views are different, the farther the output from the neck, the smaller the loss. This loss is called contrastive loss in the SimCLR paper. The loss from the head is used to adjust the ResNet and NonLinearNeck weights.

After the pre-training is completed, a downstream task is performed. The downstream task uses the ResNet learned in the pre-training as a backbone for the brain tumor detection task.

The training scheduler for the downstream task uses a cosine annealing algorithm. The learning rate is gradually attenuated using a cosine function. The learning rate is maximized immediately after the start of learning, and is minimized at the end of learning. The number of epochs is 90. LARS is used as the optimization algorithm, as in the pre-training.

The data augmentation used in the downstream task is described in Section \ref{downstreamtask-data-aug}. The downstream task model consists of three modules: ResNet, GlobalAveragePooling, and LinearClsHead. GlobalAveragePooling is the global average pooling layer, and LinearClsHead is a linear classifier that uses cross-entropy error as the loss function. The loss function is weighted by the classes as described in Section \ref{class-label-subsection}, and the ResNet network weights are adjusted by backpropagation from the losses computed by LinearClsHead.

\subsubsection{MFF}

SimCLR uses contrastive learning as its self-supervised learning method, while MFF~\cite{mffpaper} uses Masked Image Modeling (MIM).

First, the specific algorithm of MIM is explained. First, the image is divided into squares of equal size. The divided squares are called patches. Then, some patches are randomly selected and masked. That is, they are hidden. The masked patches are replaced by a special embedding called mask embedding. The patched images are then input to the model. The model performs a pretext task to reconstruct the original image from the patched image. The model then performs a pre-text task to reconstruct the original image from the patched image, i.e., inferring the masked area from the unmasked area. Whereas in contrast learning, the pretext task is to compare the two views, in MIM the pretext task is to generate the source image from the masked image.

Next, the MFF pre-training method is described. The image is divided into 16x16 patches. Since the input image is scaled down to 224x224, 196 patches are created from a single image. Then, 75\% of all patches are randomly selected and masked. The patches are then input to the model.

The backbone of the model is Vision Transformer (ViT)~\cite{vitpaper}. The model setup uses MFFViT, which is a plain ViT implementation with an additional masking process. Neck's MAEPretrainDecoder attempts to reconstruct the original image from the input. The result of the reconstruction is fed to the head, MAEPretrainHead. The module calculates the loss from the original image and the reconstruction result. The parameters of MFFViT and MAEPretrainDecoder are adjusted to minimize the loss.

For the pre-training scheduler, the learning rate was changed from the standard MMPreTrain setting. In the course of the experiment, we found that using MMPreTrain's standard settings for the learning rate caused the loss function to reach NaN at 40 to 50 epochs after the start of learning. In other words, the value of the loss function diverged. Possible causes for the divergence of the learning rate include the presence of NaN in some data in the dataset or an excessively high learning rate. We investigated these possible causes and found that the learning rate is too high. We searched for a learning rate at which the loss function does not diverge to NaN while gradually decreasing the learning rate, and found that a learning rate of 2.4e-4 or less is sufficient for stable learning without divergence. Therefore, we decided to set the learning rate to a maximum of 2.4e-4 in the pre-training.

After completing the pre-training, a downstream task was performed using ViT on the pre-trained backbone, using class weighting and data augmentation as in SimCLR. The training scheduler for the downstream task is the standard MMPreTrain configuration; MAEPretrainDecoder is not used for the downstream task. The data augmentation used in the downstream task is described in Section \ref{downstreamtask-data-aug}. Only the output of the final layer of ViT is used in the downstream task; the output of the final layer of ViT is input to the LinearClsHead module, which is a linear classifier.

\subsubsection{SparK}\label{spark-information}

Finally, we describe the SparK~\cite{sparkpaper} model, which also uses MIM for the pre-training task, but has the characteristics of network hierarchy and sparsity.

The hierarchical structure of the network means that the convolutional neural network (CNN) is stacked in multiple layers, with each layer gradually reducing the feature map. SparK uses a backbone called ConvNeXt~\cite{liu2022convnet}, which builds a hierarchical structure inside ConvNeXt by stacking multiple ConvNeXt blocks. In ConvNeXt, the resolution of the feature map is reduced by a factor of $\frac{1}{2}$ at each layer. Hierarchization has the advantage that features at various scales can be extracted from an image. While features from the final layer alone can only be extracted at a high level of abstraction, by extracting features from each of the layers at various scales, features at different levels of abstraction can be used in later networks.

"Sparsity" means that the coupling and structure of a neural network is sparse. In other words, it means that unnecessary parameters are reduced. The process of making a network sparse is called sparsification. Sparsification has the advantage of improving generalization performance in addition to speeding up processing and reducing memory usage ~\cite{hoefler2021sparsity}. Sparsification also has the advantage that mask pattern loss does not occur, as explained in section 3.1 of SparK's paper ~\cite{sparkpaper}, where it is pointed out that when a non-sparse CNN is used, the mask pattern is lost with each pass through the layers when a masked image is input. However, sparse CNNs are not as efficient as masked CNNs. However, a sparse CNN skips the calculation of the masked part, so the pattern loss does not occur and the effect of the mask can be applied to all layers.

We now describe the SparK pre-training algorithm. SparK, like MFF, uses MIM as a self-supervised learning method. Images are first divided into patches. The size of the patches is 32x32, and a total of 49 patches are generated. Of those patches, 60\% are masked and a pretext task is performed to reconstruct the masked areas.

The network consists of encoders and decoders. ConvNeXt is used as the backbone. The input image is first divided into square patches. Then, 60\% of the total patches are masked. The masked image is then input to the SparseConvNeXt module as the encoder. The SparseConvNeXt network consists of four stages, each of which contains a sparsified convolutional neural network. As the data passes through the stages, the resolution of the data decreases. Ultimately, the output from the final layer of the encoder is the feature of the masked image. SparKLightDecoder is a decoder that attempts to recover the original, unmasked image from the output of SparseConvNeXt. The decoder network, like the encoder, is hierarchical, increasing in resolution as it passes through each stage until it finally results in a reconstructed image. Based on this loss, SparseConvNeXt and SparKLightDecoder parameters are updated to reconstruct the original image from the masked image. For the downstream task, we changed MMPreTrain's learning scheduler from the standard one, because we experienced problems with loss reduction when using MMPreTrain's learning scheduler. The experiments with different hyperparameters showed that the learning rate was too high; the standard MMPreTrain setting had a maximum learning rate of 3.2e-3, which was changed to 1e-4 in this experiment. We also increased the number of epochs from 300 to 700 because we thought that a lower learning rate would require more time for learning.

The downstream task model does not use the decoder, SparKLightDecoder, but only the encoder, ConvNeXt. The output of the backbone is fed into LinearClsHead, a linear classifier, which performs the class classification; as with SimCLR and MFF, LinearClsHead performs the class weighting.

\begin{table}[H]
 \caption{Results of SimCLR}
 \label{table:simclr_result}
 \centering
  \begin{tabular}{clll}
   \hline \hline
   Augmentation & Precision & Recall & F1-Score \\
   \hline
        None & 55.6 & 96.2 & 70.4\\
        RandomResizedCrop & 52.2 & 92.3 & 66.7 \\
        RandomFlip & 53.2 & 96.2 & 68.5 \\
        CutMix & 49.0 & 92.3 & 64.0 \\
        Mixup & 62.9 & 84.6 & \bf{72.1} \\
   \hline
  \end{tabular}
\end{table}

\begin{table}[H]
\caption{Results of MFF}
 \label{table:mff_result}

 \centering
  \begin{tabular}{clll}
   \hline \hline
   Augmentation & Precision & Recall & F1-Score \\
   \hline
        None & 82.6 & 73.1 & \bf{77.6} \\
        RandomResizedCrop & 75.0 & 69.2 & 72.0 \\
        RandomFlip & 76.0 & 73.1 & 74.5 \\
        CutMix & 56.4 & 84.6 & 67.7 \\
        Mixup & 66.7 & 69.2 & 67.9 \\
   \hline
  \end{tabular}
\end{table}

\begin{table}[H]
\caption{Results of SparK}
 \label{table:spark_result}
 \centering
  \begin{tabular}{clll}
   \hline \hline
   Augmentation & Precision & Recall & F1-Score \\
   \hline
        None & 31.3 & 19.2 & 23.8 \\
        RandomResizedCrop & 80.6 & 96.2 & \bf{87.7} \\
        RandomFlip & 55.6 & 19.2 & 28.6 \\
        CutMix & 41.9 & 69.2 & 52.2 \\
        Mixup & 13.0 & 23.1 & 16.7 \\
   \hline
  \end{tabular}
\end{table}
\subsubsection{Results}\label{basic-information-result}

\begin{figure}[H]
\centering
\includegraphics[scale=0.35]{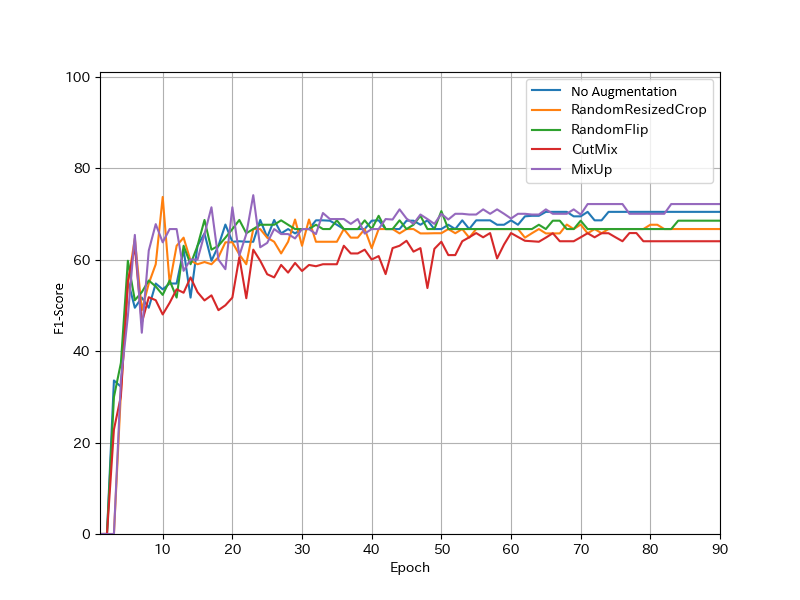}
\caption{F1 Scores for SimCLR}
\label{fig:simclr-f1-trans}
\end{figure}

\begin{figure}[H]
\centering
\includegraphics[scale=0.35]{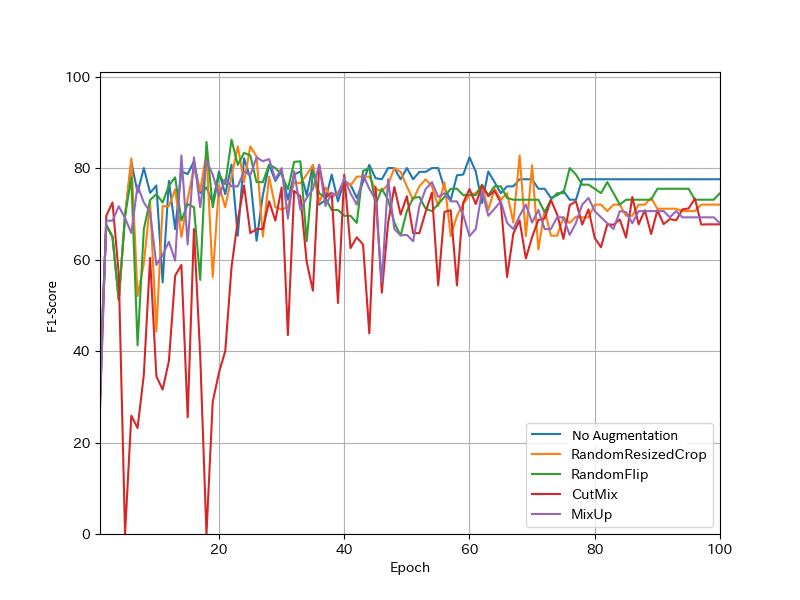}
\caption{F1 Scores for MFF}
\label{fig:mff-f1-trans}
\end{figure}

\begin{figure}[H]
\centering
\includegraphics[scale=0.35]{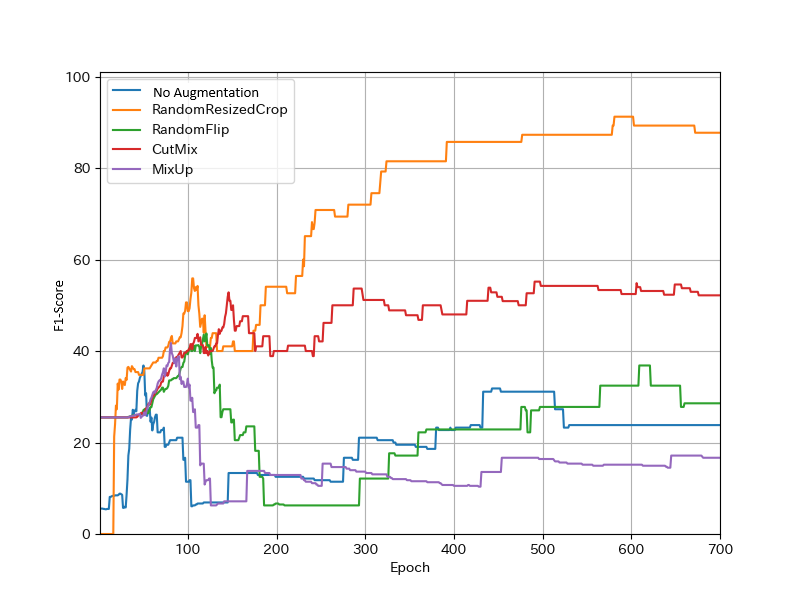}
\caption{F1 Scores for SparK}
\label{fig:spark-f1-trans}
\end{figure}

Tables \ref{table:simclr_result}, \ref{table:mff_result}, and \ref{table:spark_result} are the results of training the SimCLR, MFF, and Spark downstream tasks to their final epochs. Figures \ref{fig:simclr-f1-trans}, \ref{fig:mff-f1-trans}, and \ref{fig:spark-f1-trans} are the F1 scores for the downstream tasks of SimCLR, MFF, and SparK, respectively. SparK achieved an F1 score of 87.7 when using RandomResizedCrop, the highest among the three self-supervised learning methods. The recall rate reached 96.2\%, the same as the highest value of the other methods. In contrast, the fit rate was only 80.6\%. 82.6\% was achieved when the highest F1 score was achieved with MFF, so SparK tends to have slightly higher false negatives when RandomResizedCrop is used. However, in the brain tumor detection task, the failure to detect a brain tumor is more problematic than a false positive, so reproducibility is preferred over goodness-of-fit. Therefore, SparK, which has a somewhat high fit rate and the highest reproducibility under the conditions using RandomResizedCrop, is adopted.

However, SparK performs well when using RandomResizedCrop, but performs worse than the other models when using the other data augmentation methods. This means that SparK is more affected by data augmentation than the other two models. Therefore, the choice of data augmentation method is very important for SparK.

\vspace{-0.3cm}

\vspace{-0.3cm}

\vspace{-0.3cm}

\section{Method}
\label{OurMethod}


In the self-supervised learning method known as the Masked Image Modeling (MIM), an image is masked by some algorithm, and the model is trained through the pretext task of reconstructing the original image from the masked one. Therefore, the choice of masking pattern holds significant meaning, and it is expected that changing the masking pattern will also change how the model extracts features from the image. In fact, in SimMIM~\cite{xie2022simmim}, differences in masking patterns led to variations in performance. Therefore, we decided to investigate whether altering the algorithm for determining the masked regions could enhance the feature extraction ability of the backbone.

\begin{figure}[h]
    \begin{tabular}{cc}
        \begin{minipage}[t]{0.48\linewidth}
            \centering
            \includegraphics[width=0.9\linewidth]{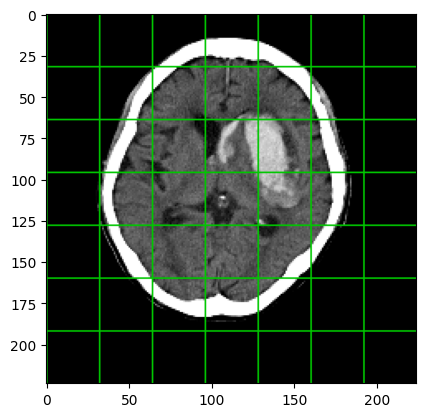}
            \caption{Before Masking Application}
            \label{unmasked-raw-image}
        \end{minipage} &
        \begin{minipage}[t]{0.48\linewidth}
            \centering
            \includegraphics[width=0.9\linewidth]{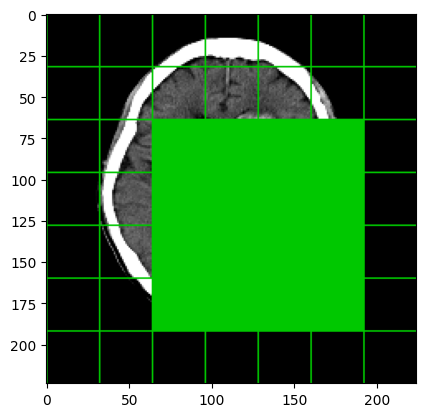}
            \caption{Square Mask}            
            \label{square-masked-image}
        \end{minipage} \\

        \begin{minipage}[t]{0.48\linewidth}
            \centering
            \includegraphics[width=0.9\linewidth]{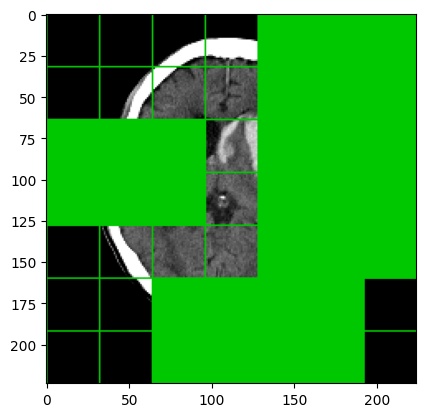}
            \caption{Block-wise Mask}
            \label{block-wise-masked-image}
        \end{minipage} &
         \begin{minipage}[t]{0.48\linewidth}
            \centering
            \includegraphics[width=0.9\linewidth]{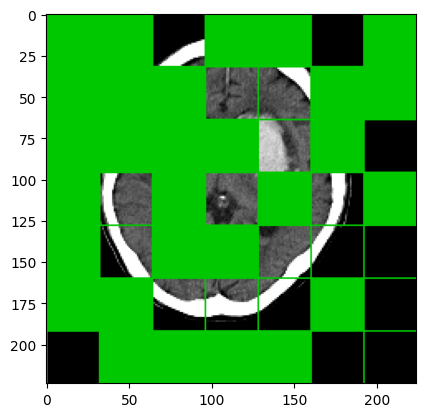}
            \caption{Random Mask}            
            \label{random-masked-image}
        \end{minipage} \\

        \begin{minipage}[t]{0.48\linewidth}
            \centering
            \includegraphics[width=0.9\linewidth]{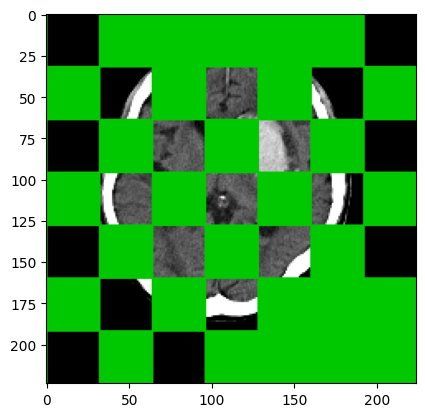}
          \caption{Mesh Mask(60\%)}
          \label{check-masked-60-image}
        \end{minipage} &
        \begin{minipage}[t]{0.48\linewidth}
            \centering
            \includegraphics[width=0.9\linewidth]{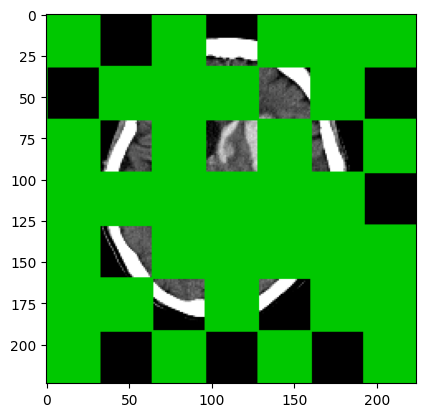}
            \caption{Mesh Mask(70\%)}            
            \label{check-masked-70-image}
        \end{minipage}
    \end{tabular}
\end{figure}

In this paper, we propose a new masking pattern called the mesh mask. As seen in Figure
\ref{unmasked-raw-image}, 
brain tumors are located in small regions of the brain. In SimMIM, square masks (Figure \ref{square-masked-image}),
block-wise masks 
(Figure \ref{block-wise-masked-image}), 
and random masks (Figure \ref{random-masked-image})
were used, but these patterns may hide the entire tumor. For instance, in Figure \ref{square-masked-image},
the square mask is used, but the brain tumor (the white region inside the brain) seen in Figure
\ref{unmasked-raw-image} 
is completely hidden. In such cases, the model is required to reconstruct the tumor without any trace of it, which may hinder the model's ability to effectively learn tumor features. As shown in Section 4.4 of the SimMIM paper, when an object in an image is completely masked, effectively erasing all traces of the object, the model reconstructs the image assuming that the object does not exist. Therefore, by applying a mesh-like mask, as shown in Figure \ref{check-masked-70-image},
and ensuring that part of the tumor remains exposed, we aim to actively encourage the reconstruction of the tumor, potentially improving the model's ability to extract brain tumor features.
\begin{figure}[H]
  \begin{minipage}[b]{0.49\linewidth}
    \centering
    \includegraphics[width=0.9\linewidth]{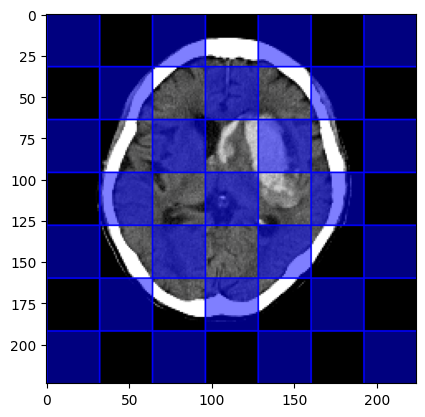}
  \end{minipage}
 \begin{minipage}[b]{0.49\linewidth}
    \centering
    \includegraphics[width=0.9\linewidth]{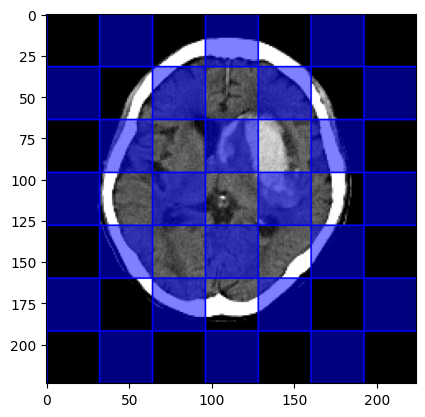}
  \end{minipage}
  \caption{Two possible candidates for non-masked patches used in the mesh mask. The blue semi-transparent patches represent the candidates for non-masked patches.}
  \label{nonmask-patch-selection}
\end{figure}

We now describe the specific algorithm for the mesh masking technique. The overview of this algorithm is outlined in Algorithm \ref{alg-mesh-mask-algo}.
\begin{algorithm}[H]
\caption{Algorithm for the Mesh Mask}
\label{alg-mesh-mask-algo}
\begin{algorithmic}[1]

\Function {generate\_mesh\_mask}{mask\_ratio}
    \If  {$ \Call{Rand}{0, 1} > 0.5 $}
        \State {$ K \gets \{(i, j) \mid i, j \in \{0, 1, 2, \cdots, 7\}, 7j + i \equiv 0 \mod 2 \} $}
    \Else
        \State {$ K \gets \{(i, j) \mid i, j \in \{0, 1, 2, \cdots, 7\}, 7j + i \equiv 1 \mod 2 \} $}
    \EndIf
    \State {$ num\_non\_masked\_patches \gets \Call{Round}{(1 - mask\_ratio)(7 \cdot 7)} $}
    \State {$ L \gets \Call{Shuffle}{K}[:num\_non\_masked\_patches] $}
    \State {$ M \gets \{(i,j) \mid i, j \in \{0, 1, 2, \cdots, 7\}\} - L $}
    \State \Return {$ M $}
\EndFunction

\end{algorithmic}
\end{algorithm}
First, we prepare the candidate non-masked patches, i.e., the patches that will not be masked. The candidates for the non-masked patches consist of two possible mesh structures, as shown in Figure \ref{nonmask-patch-selection}, and these structures are selected randomly with a probability of 50\%. We now express these candidate patches mathematically. The image is divided into patches, with a size of 7×7. Therefore, the set of coordinates for the non-masked patch candidates is:
\begin{equation}
K = \{(i, j) \mid i, j \in \{0, 1, 2, \cdots, 7\}, 7j + i \equiv 0 \mod 2 \}
\end{equation}
or
\begin{equation}
K = \{(i, j) \mid i, j \in \{0, 1, 2, \cdots, 7\}, 7j + i \equiv 1 \mod 2 \}
\end{equation}
This set represents the candidate non-masked patches, where the coordinate of the top-left patch is designated as (0, 0). Next, we select the actual non-masked patches from the set K. The number of patches to be selected as masked patches is determined by the mask ratio.

The masking ratio refers to the proportion of masked patches relative to the total number of patches. Let the mask ratio be denoted as n; the proportion of non-masked patches is then 1 - n. Therefore, a total of $\lfloor(1 - n)(7 \cdot 7)\rfloor$ non-masked patches are randomly selected from set K. Since the mask ratio is always set to be at least 50\%, the number of non-masked patches selected is always $|K|$ or fewer. The set of coordinates for the selected non-masked patches is denoted as L. Consequently, the set of masked patches M is defined as:
$M = \{(i,j) \mid i, j \in \{0, 1, 2, \cdots, 7\}\} - L$.
Figure \ref{check-masked-60-image}
shows an example of a masking pattern with a 60\% mask ratio, and Figure \ref{check-masked-70-image}
shows an example with a 70\% mask ratio.

\section{Experiment}







The experimental setup employs MMPreTrain, a framework designed for self-supervised learning. The configuration used in MMPreTrain is the same as described in Section \ref{spark-information}. The self-supervised learning model used is SparK, and for data augmentation in downstream tasks, we adopt only the RandomResizedCrop method. An overview of SparK is provided in Section \ref{spark-information}, while data augmentation strategies for downstream tasks are described in Section \ref{downstreamtask-data-aug}.

For the pretraining of SparK, we adopt a self-supervised learning approach known as Masked Image Modeling (MIM). In this experiment, we investigate how different masking patterns used in MIM—namely, square mask, block-wise mask, random mask, and mesh mask—affect downstream task performance.

The square mask applies square-shaped masks to random locations in the image, with a fixed mask size. In SimMIM~\cite{xie2022simmim}, the highest accuracy of 82.6\% was achieved when using a 2×2 mask. An example of a 4×4 square mask applied to an image is shown in Figure \ref{square-masked-image}. In this study, we follow the settings of the SimMIM paper and conduct experiments using 2×2, 3×3, and 4×4 square masks.

The block-wise mask, proposed in BEiT~\cite{mimpaper}, is a more complex masking strategy. This method applies multiple rectangular masks of random sizes until a predefined masking ratio is reached. Figure 
\ref{block-wise-masked-image} 
illustrates an example where the masking ratio is 60\%. Following the SimMIM paper, we experiment with masking ratios of 40\%, 60\%, and 80\%.

The random mask is the default masking strategy used in the original SparK paper~\cite{sparkpaper}. SimMIM~\cite{xie2022simmim} reported that this method exhibited stable performance compared to other masking strategies. In this approach, input images are first divided into patches. In our experiments, the patch size is 32×32 and the input image size is 224×224, resulting in a total of 49 patches. This unmasked state is illustrated in Figure 
\ref{unmasked-raw-image}. 
A fixed proportion of these 49 patches is then randomly selected and masked. We test masking ratios of 50\%, 60\%, and 70\%, consistent with the 60\% ratio used in the SparK paper. Figure \ref{random-masked-image} shows an example of a 60\% mask ratio, in which 29 out of 49 patches are masked.

For the mesh mask, we conduct experiments with masking ratios of 50\%, 60\%, 70\%, and 80\%. Details on the mesh masking method can be found in Section \ref{OurMethod}.

\section{Results}






The results are summarized in Tables \ref{table:square-mask-result}, \ref{table:blockwise-mask-result}, \ref{table:random-mask-result}, and \ref{table:check-mask-result}. For the square mask, the highest F1 score of 83.6 was achieved when using a 2×2 mask size. In the case of the block-wise mask, the F1 score remained consistent across different masking ratios, with a score of 83.2. For the random mask, the highest F1 score of 87.7 was observed at both 50\% and 60\% masking ratios. In the case of the mesh mask, the best performance was achieved at a 70\% masking ratio, also yielding an F1 score of 87.7.

In the SimMIM paper, it was reported that the performance of the random masking method tends to degrade as the masking ratio deviates from a certain optimal value. However, in our study, the mesh mask exhibited the best performance at a 70\% masking ratio, followed by 50\%, with the lowest performance at 60\%. This unexpected trend may be attributed to the characteristics introduced by the meshing pattern of the mask, although the underlying mechanism remains unclear.

While the random mask achieved its best F1 score near a 50\% masking ratio, the mesh mask performed best at a higher masking ratio of 70\%. This may be due to the fact that mesh masking tends to distribute masked patches more evenly across the entire image compared to random masking.

Compared to block-based masking methods such as square and block-wise masks, which obscure multiple patches together, patch-level masking approaches such as random and mesh masks resulted in F1 scores that were over 5\% higher. This suggests that in the context of pretraining SparK on brain imaging data, the choice of masking pattern plays a critical role, and patch-level masking is particularly effective.

Although the mesh mask outperformed the square and block-wise masks, it showed performance comparable to that of the random mask and did not surpass it. This similarity may be due to the fact that both mesh and random masking apply masking at the patch level, and the mesh mask pattern is structurally similar to that of the random mask, potentially explaining the lack of significant performance difference.

\begin{table}[H]
\caption{Square Mask}
 \label{table:square-mask-result}
 \centering
  \begin{tabular}{cllll}
   \hline \hline
   Masking Type & Accuracy & Precision & Recall & F1 Score \\
   \hline
        2×2 & 94.9 & 79.3 & 88.5 & \bf{83.6} \\
        3×3 & 94.4 & 76.7 & 88.5 & 82.1  \\
        4×4 & 94.4 & 76.7 & 88.5 & 82.1  \\
   \hline
  \end{tabular}
\end{table}

\begin{table}[H]
\caption{Block-wise Mask}
 \label{table:blockwise-mask-result}
 \centering
  \begin{tabular}{cllll}
   \hline \hline
    Masking Ratio & Accuracy & Precision & Recall & F1 Score \\  
   \hline
       40\% & 94.4 & 75.0 & 92.3 &  \bf{82.8} \\
       60\% & 94.4 & 75.0 & 92.3 &  \bf{82.8} \\
       80\% & 94.4 & 75.0 & 92.3 &  \bf{82.8} \\
   \hline
  \end{tabular}
\end{table}

\begin{table}[H]
\caption{Random Mask}
 \label{table:random-mask-result}
 \centering
  \begin{tabular}{cllll}
   \hline \hline
     Masking Ratio & Accuracy & Precision & Recall & F1 Score \\  
   \hline
        50\% & 96.1 & 80.6 & 96.2 & \bf{87.7} \\
        60\% & 96.1 & 80.6 & 96.2 & \bf{87.7} \\
        70\% & 94.9 & 77.4 & 92.3 & 84.2 \\
   \hline
  \end{tabular}
\end{table}

\begin{table}[H]
\caption{Mesh Mask}
 \label{table:check-mask-result}
 \centering
  \begin{tabular}{cllll}
   \hline \hline
     Masking Ratio & Accuracy & Precision & Recall & F1 Score \\ 
   \hline
        50\% & 95.5 & 78.1 & 96.2 & 86.2 \\
        60\% & 94.9 & 77.4 & 92.3 & 84.2 \\
        70\% & 96.1 & 80.6 & 96.2 & \bf{87.7} \\
        80\% & 95.5 & 80.0 & 92.3 & 85.7 \\
   \hline
  \end{tabular}
\end{table}

\section{Conclusions}\label{結論}



In this study, we compared the SparK model, which incorporates sparsity and hierarchical structure, with other self-supervised learning models, and confirmed the superiority of this architecture. Furthermore, we found that the masking pattern used during SparK pretraining has a non-negligible impact on downstream task performance. In particular, patch-level masking methods such as the random mask and mesh mask demonstrated better performance than block-level masking methods such as the square mask and block-wise mask.

There are several directions for future work. First, further investigation is needed into the impact of masking patterns on model performance. Although the mesh mask proposed in this study outperformed the square and block-wise masks, its performance was comparable to that of the random mask. Given the critical role of masking strategies in masked image modeling, it is important to conduct more in-depth studies to better understand their influence.

Second, the development of new masking patterns that can achieve higher performance is a key challenge. In this study, none of the evaluated masking strategies surpassed the performance of the existing random mask. Thus, additional research is needed to design and evaluate novel masking algorithms that can further improve model performance.

\bibliography{refe}

\appendix
\section{GradCam Images of Spark Models With RandomResizedCrop Applied}

\begin{figure}
\centering
  \begin{minipage}[b]{0.45\linewidth}
    \centering
    \includegraphics[width=\linewidth]{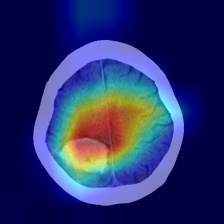}
  \end{minipage}
  \begin{minipage}[b]{0.45\linewidth}
    \centering
    \includegraphics[width=\linewidth]{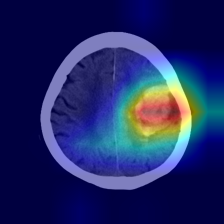}
  \end{minipage}
  \caption{Example Grad-CAM visualizations for SparK model on positive test images}
  \label{spark-gradcam}
\end{figure}

Figure \ref{spark-gradcam} presents the Grad-CAM~\cite{gradcam-paper} visualization for a positively labeled test image, generated using a SparK model pretrained with RandomResizedCrop. This visualization highlights the regions of the image that the model considered as important when making its positive classification. The white mass inside the brain represents the tumor. The red regions indicate areas to which the model responded strongly. As the red highlights appear over or near the tumor region, it can be inferred that the model made its positive prediction based on the presence of the tumor.

\section{Difference in Performance of SparK Model with and without Sparsity}


We also evaluated the performance of a non-sparse variant of SparK. Table \ref{table:normal_spark_and_nonsparse_result} compares the standard SparK model with one that uses a non-sparse ConvNeXt backbone. In both cases, RandomResizedCrop was employed as the data augmentation method, and all conditions were kept identical except for the presence or absence of sparsity.
While recall remained the same across both models, the non-sparse SparK showed an 11.2\% decrease in precision. As described in Section \ref{spark-information}, the incorporation of sparsity contributes to improved generalization performance and helps prevent the masking pattern from vanishing during masked image modeling (MIM) pretraining. These results confirm that sparsity indeed contributes to the enhanced performance of the SparK model.

\begin{table}[H]
\caption{Performance comparison of SparK models with and without sparsity}
 \label{table:normal_spark_and_nonsparse_result}
 \centering
  \begin{tabular}{clll}
   \hline \hline
   Model & Precision & Recall & F1 Score \\
   \hline
    Sparse SparK & 80.6 & 96.2 & \textbf{87.7} \\
Non-sparse SparK & 69.4 & 96.2 & 80.6 \\    
   \hline
  \end{tabular}
\end{table}

\end{document}